\documentclass[a4paper]{spie} 

\usepackage{amsmath,amsfonts,amssymb}
\usepackage{graphicx}
\usepackage[colorlinks=true, allcolors=blue]{hyperref}
\usepackage{subcaption}
\usepackage{aas_macros}
\title{SIOUX project: a simultaneous multiband camera for exoplanet atmospheres studies}

\author[a,b]{Jean Marc Christille}
\author[b]{Aldo Stefano Bonomo}
\author[c]{Francesco Borsa}
\author[b]{Deborah Busonero}
\author[a]{Paolo Calcidese}
\author[f]{Riccardo Claudi}
\author[b]{Mario Damasso}
\author[b]{Paolo Giacobbe}
\author[d,e]{Emilio Molinari}
\author[g]{Emanuele Pace}
\author[b]{Alberto Riva}
\author[b]{Alessandro Sozzetti}
\author[e]{Giorgio Toso}
\author[e]{Daniela Tresoldi}

\affil[a]{Astronomical Observatory of the Autonomous Region of the Aosta Valley (OAVdA), 
Loc. Lignan 39, I-11020, Nus (AO), Italy}
\affil[b]{INAF - Osservatorio Astrofisico di Torino, Strada Osservatorio 20, I-10025, Pino Torinese (TO), 
Italy}
\affil[c]{INAF - Osservatorio Astronomico di Brera, Via Emilio Bianchi 46, I-23807 Merate (LC), Italy}
\affil[d]{FGG-INAF, Telescopio Nazionale Galileo, Rambla J.A. Fernández P. 7, E-38712 Bre\~{n}a Baja, TF - 
Spain}
\affil[e]{INAF - IASF Milano, via Edoardo Bassini 15, I-20133, Milano (MI), Italy}
\affil[f]{INAF - Osservatorio Astronomico di Padova, Vicolo Osservatorio 5 - 35122 - Padova (PD), Italy}
\affil[g]{Universit\'a degli Studi di Firenze, Dipartimento di Fisica e Astronomia, Via Giovanni Sansone 1,50019 Sesto  Fiorentino, Italy}

\authorinfo{Further author information: (Send correspondence to Jean Marc Christille)\\Jean Marc Christille: E-
mail: jeanmarc.christille@gmail.com, Telephone: 00393294926742}

\begin{document} 
\maketitle

\begin{abstract}

The exoplanet revolution is well underway. 
The last decade has seen order-of-magnitude increases in the number of known planets beyond the Solar system. 
Detailed characterization of exoplanetary atmospheres provide the best means for distinguishing the makeup of 
their outer layers, and the only hope for understanding the interplay between initial composition chemistry, temperature-pressure atmospheric profiles,
dynamics and circulation. 

While pioneering work on the observational side has produced the first important detections of atmospheric 
molecules for the class of transiting exoplanets, important limitations are still present due to the lack of 
systematic, repeated measurements with optimized instrumentation at both visible (VIS) and near-infrared (NIR) 
wavelengths. 
It is thus of fundamental importance to explore quantitatively possible avenues for improvements. 
In this paper we report initial results of a feasibility study for the prototype of a versatile multi-band 
imaging system for very high-precision differential photometry that exploits the choice of specifically 
selected narrow-band filters and novel ideas for the execution of simultaneous VIS and NIR measurements.

Starting from the fundamental system requirements driven by the science case at hand, we describe a set of 
three opto-mechanical solutions for the instrument prototype: 1) a radial distribution of the optical flux 
using dichroic filters for the wavelength separation and narrow-band filters or liquid crystal filters for the 
observations; 2) a tree distribution of the optical flux (implying 2 separate foci), with the same technique 
used for the beam separation and filtering; 3) an 'exotic' solution consisting of the study of a complete 
optical system (i.e. a brand new telescope) that exploits the chromatic errors of a reflecting surface for 
directing the different wavelengths at different foci.

In this paper we present the first results of the study phase for the three solutions, as well as the results 
of two laboratory prototypes (related to the first two options), that simulate the most critical aspects of the 
future instrument. 

\end{abstract}


\newpage%

\section{Scientific Rationale: The Atlas of Exoplanetary Atmospheres}

The exoplanet revolution is well underway. More than 3\,000 extrasolar planets are known today and 
those which have been well characterised show an astonishing diversity concerning 
their orbital (period, semi-major axis, eccentricity) and physical (radius, mass, density) 
parameters; this diversity is related to different formation and evolution histories. 

The majority of the discovered exoplanets are small-size and low-mass planets. Indeed, planet occurrence rates derived from 
both radial velocity surveys and the \emph{Kepler} space telescope indicate that small and low-mass planets, 
i.e. mini-Neptunes, super-Earths and Earth-sized planets, occur far more frequently than giant planets 
(e.g., Howard et al. 2010\cite{2010Sci...330..653H}~; Fressin et al. 2013\cite{2013ApJ...766...81F}), 
and they are two-three times more common around M dwarfs than FGK main-sequence stars 
(Mulders et al. 2015\cite{2015ApJ...814..130M}).
In particular, more than 50\% of M dwarfs have at least one planet with $1 \le R_{\rm p} \le 2~\rm R_{\oplus}$ 
and orbital period $P < 50$~d according to Dressing \& Charbonneau (2015)\cite{2015ApJ...807...45D}~. 
The frequency of giant planets orbiting solar-like stars is $\sim 5\%$ for $P < 400$~d and 
decreases to $\sim 1\%$ for $P < 10$~d (e.g., Santerne et al. 2016\cite{2016A&A...587A..64S}), 
that is for the so-called hot Jupiters (with high equilibrium temperatures, $T_{\rm eq} \gtrsim 1000$~K, 
given their proximity to the host star).

The atmospheres of giant planets can be investigated with different techniques such as 
i) spectrophotometry and low-dispersion spectroscopy  
of transits and secondary eclipses from UV to mid-IR, 
ii) NIR and mid-IR phase curves, 
iii) high-dispersion spectroscopy, and 
iv) direct imaging 
(see, e.g., the reviews by Seager \& Deming 2010\cite{2010ARA&A..48..631S}~, 
Burrows 2014\cite{2014Natur.513..345B} and Crossfield 2015\cite{2015PASP..127..941C}~, and references therein). 
In particular, transit spectrophotometry consists in measuring the transit depth (hence the planetary radius)
at multiple wavelengths ($\lambda$) to find out at which wavelengths the transit is deeper ($R_{\rm p}$ is larger) 
and thus the atmosphere is more opaque because of atomic and/or molecular transitions. 
In such a way, it is possible to obtain the planet's ``transmission spectrum". 
A theoretical transmission spectrum computed by Fortney et al. (2010)\cite{2010ApJ...709.1396F} is shown in Fig.~\ref{fig:transitfit} (dotted line) for a hot Jupiter with 
$R_{\rm p}=1.3~R_{\rm Jup}$, $T_{\rm eq} = 2000$~K, and surface gravity $g=25~\rm m~ s^{-2}$; the planetary radius is larger at the wavelength of the sodium and potassium doublets because
of absorption of stellar light by these alkali species in the upper planetary atmosphere.

Transmission spectra have been obtained with spectrophotometry for more than twenty hot Jupiters orbiting relatively 
bright host stars ($V < 12.5$) with both space-born facilities mainly onboard the Hubble Space Telescope (HST) and 
ground-based instrumentation at telescopes with apertures generally larger than 3.5~m 
(WHT, VLT, GTC, Gemini-South, etc.). Planets with ``inflated" radii (e.g., Baraffe et al. 2014\cite{2014prpl.conf..763B}~) 
and high temperatures are the most promising targets for atmospheric characterization 
thanks to their relatively large atmospheric scale heights (defined by $H = kT /\mu_m g$,
where $k$ is Boltzmann's constant, $T$ is temperature, $\mu_m$ is mean molecular weight, 
and $g$ is surface gravity). 

The results obtained up to now show a surprising diversity of the atmospheres of hot Jupiters.  
Both Na and K have been detected for several of them but in some cases only one of the two species 
(Na or K) has been found. Some transmission spectra are ``clear" with possibly distinct pressure-broadened wings 
of Na and K features (Fischer et al. 2016\cite{2016arXiv160104761F}~); 
other spectra are dominated by hazes, i.e. clouds composed by small-size particles ($< 0.1~\rm \mu m$) that attenuate the Na and K absorption features and give rise to stronger Rayleigh scattering than clear spectra
(e.g., Pont et al. 2013\cite{2013MNRAS.432.2917P}~, Nikolov et al. 2015\cite{2015MNRAS.447..463N}~); 
clouds with larger particles may flatten to a greater extent the planetary spectra especially in the NIR 
(Sing et al. 2015\cite{2015MNRAS.446.2428S}~).
Moreover, recent observations with the WFC3 on board the HST 
led to the detection of H$_2$O in the atmosphere of some hot Jupiters 
but with a variety of absorption amplitudes (e.g., Kreidberg et al. 2014a\cite{2014ApJ...793L..27K}~, 
Madhusudhan et al. 2014\cite{2014ApJ...791L...9M}~).
These H$_2$O varying amplitudes seem to be caused by different levels of obscuration by hazes/clouds
rather than primordial water depletion during planet formation (Sing et al. 2016\cite{2016Natur.529...59S}~).
However, the composition of these hazes/clouds and the reason why they are present at high 
altitudes in some exoplanetary atmospheres but not in others are completely unknown.

Detections of Na and H$_2$O have been used to determine vertical temperature-pressure 
profiles (e.g., Sing et al. 2008\cite{2008ApJ...686..667S}~, Stevenson et al. 2014\cite{2014Sci...346..838S}~). 
In addition to Na, K, and H$_2$O, also TiO and VO are expected to be found in the atmospheres of the hottest giant planets 
(Fortney et al. 2008\cite{2008ApJ...678.1419F}~) and their presence would leave clear imprints in the 
transmission optical spectrum but no firm detection has
been reported up to now. These species are thought to produce atmospheric temperature inversions 
according to theoretical models (even though several other compounds have also been suggested), 
and this prediction might be tested with emission spectroscopy in the near- and mid-IR, 
in case of TiO/VO detection.

Transmission spectra of a few smaller planets, such as Neptune-size planets or mini-Neptunes, have been obtained
with spectrophotometry and turned out to be flat, i.e. they do not show any molecular features. 
For the time being, it is not clear whether such flat spectra are due to high mean molecular weights 
(H-poor atmospheres) or the presence of hazes/clouds (e.g., Kreidberg et al. 2014b\cite{2014Natur.505...69K}~, 
Knutson et al. 2014\cite{2014Natur.505...66K}~). 

Here we report initial results of a feasibility study for the prototype of a versatile multi-band imaging 
system that exploits the choice of specifically selected narrow-band filters and novel ideas for the execution 
of simultaneous measurements at VIS \& NIR\footnote{according to the specifications for the NIR observations, sites like Paranal and Dome C (Antarctica) have atmospheric conditions transparent enough at the considered NIR bands (Sect.~\ref{subsec:OptoMechanicalSolutions}).} wavelengths. 
This instrument will allow us to obtain planetary transmission spectra from $0.35$ to $1.7~\rm \mu m$ and thus 
search at the same time for Na, K, TiO/VO, and H$_2$O   
as well as the presence of hazes and/or clouds, even in a single observing night in some cases\footnote{depending also on the 
aperture of the telescope on which it will be possibly mounted.}. 
Therefore it would represent a step forwards with respect to current instrumentation 
given that multiple observations, sometimes separated by years, with multiple instruments are nowadays required to obtain 
a transmission spectrum in the same wavelength range. 
Moreover, simultaneous observations with our prototype will allow us to better 
correct for nightly (atmospheric and instrumental) variations and for long-term
stellar behaviour due to changing starspot coverages, 
which may affect the derived planetary radius (e.g., Ballerini et al. 2012\cite{2012A&A...539A.140B}~). 


Only the GROND instrument mounted on the 2.2~m MPI/ESO telescope 
is capable to-date to undertake truly simultaneous observations at both optical and NIR wavelengths 
(Greiner et al. 2008\cite{2008PASP..120..405G}). 
However, the use of standard broad-band filters hampers the interpretation of any radius variation with wavelength (see, e.g, Mancini et al. 2013\cite{2013MNRAS.430.2932M}~, 2014\cite{2014MNRAS.443.2391M}), as most of the effects related to the presence of specific atmospheric compounds are materialized 
in terms of sharp spikes in $R_{\rm p}(\lambda)$ with widths of 10 nm, or less. 
Our prototype will be able to carry out simultaneous photometric measurements with ``higher resolution", realized in this case through the 
adoption of a significant number of {\it ad-hoc} narrow-band filters.

\section{Technical Description}
\label{sec:TechnicalDescription}

We describe here the two-tiered approach established for the above mentioned feasibility study: a) the 
development of a scientific simulator aimed at quantifying the performance in the retrieval of the fundamental 
physical quantity (planetary radius vs. $\lambda$) as a function of instrument parameters and observing strategy, to establish the fundamental system requirements driving b) the investigation 
of a number of opto-mechanical solutions for the prototype.

\subsection{Scientific Simulator}
\label{subsec:ScientificSimulator}

\subsubsection{Instrument+Site Model Simulator}
\label{subsubsec:InstrumentSiteModelSimulator}

In the study of the performance of the instrument based on simulations of photometric measurements, this 
component concerns the estimate of the expected total photometric error budget for each measurement in each of the narrow passbands selected for the prototype (Sect.~\ref{subsec:OptoMechanicalSolutions}).
The calculation takes into account the Spectral Energy Distribution of the host star, the general characteristics of the optical system (telescope + filters + detector), the typical properties of a good observing site, and a possible 
observational strategy.

The parameters considered for the optical system are 

\begin{itemize}
    \item[a)] the telescope diameter;
    \item[b)] the transmission coefficients for mirrors, filters, dichroics, and other optical components. 
\end{itemize}

The parameters considered for the detectors are as follows:

\begin{itemize}
    \item[a)] quantum efficiency;
    \item[b)] readout noise per pixel;
    \item[c)] dark current per pixel;
    \item[d)] gain factor (electrons/ADU).
\end{itemize}

A pixel scale of 0.3 arcsec/pixel was considered at this step as a preliminary value, being this a reliable value for a targeted photometric follow-up. 
Finally, the parameters considered for the observing site are 
\begin{itemize}
    \item[a)] the altitude;
    \item[b)] the sky brightness (mag/arcsec$^2$), as measured at the TNG site in standard photometric bands in new 
    Moon conditions\footnote{$http://www.tng.iac.es/info/la\_palma\_sky.html$}. The adopted values must be 
    considered as upper limits for our working bands, which cover ranges smaller than those actually used for 
    the sky brightness measurements;
    \item[c)] the atmospheric transmission coefficient,  interpolated at the given passband.
\end{itemize}

The simulation assumes a constant airmass and, following one of the most innovative features of the prototype, 
different exposure times are considered for each band of interest. 
It is assumed that scientific observations are performed with the telescope slightly out of focus, as is quite 
common practice when carrying out ground-based transit photometry (e.g., Southworth et al. 2009 \cite{2009MNRAS.396.1023S}). 
This implies that the simulated PSF FWHM is nearly an order of magnitude larger than the pixel scale. 
The circular aperture to perform target photometry is an additional input parameter to the code. 
The aperture radius is set to 2.5-3 times the value of the FWHM to ensure that almost all of the incident flux 
falls within the aperture. 

The photometric error budget includes contributions from: 

\begin{itemize}
    \item[a)] readout noise and dark current;
    \item[b)] scintillation;
    \item[c)] sky background;
    \item[d)] photon noise.
\end{itemize}

For our preliminary simulations we consider a G2V star and assume three typical apparent magnitudes (V=10, 11, and 12) for the exoplanet host star. The stellar flux is calculated from the flux per unit wavelength incident upon the surface of the Earth's atmosphere (in Watts m$^{-2}$ nm$^{-1}$), which is derived from real measurements of the Hubble Space Telescope\footnote{$http://www.stsci.edu/hst/observatory/crds/calspec.html$}(HST). Specifically, we selected HD 205905 (V=6.74) as the standard star for deriving the out-of-atmosphere fluxes in our bands, which were then adjusted to the simulated stellar magnitudes. 
These flux values are then multiplied by the collecting area of the telescope and the exposure times, and are 
scaled taking the transmission coefficients of the optics and the atmosphere into account, to simulate the number 
of photons reaching the detector. 

\begin{figure}[h!]
\begin{center}
\includegraphics[scale=0.65]{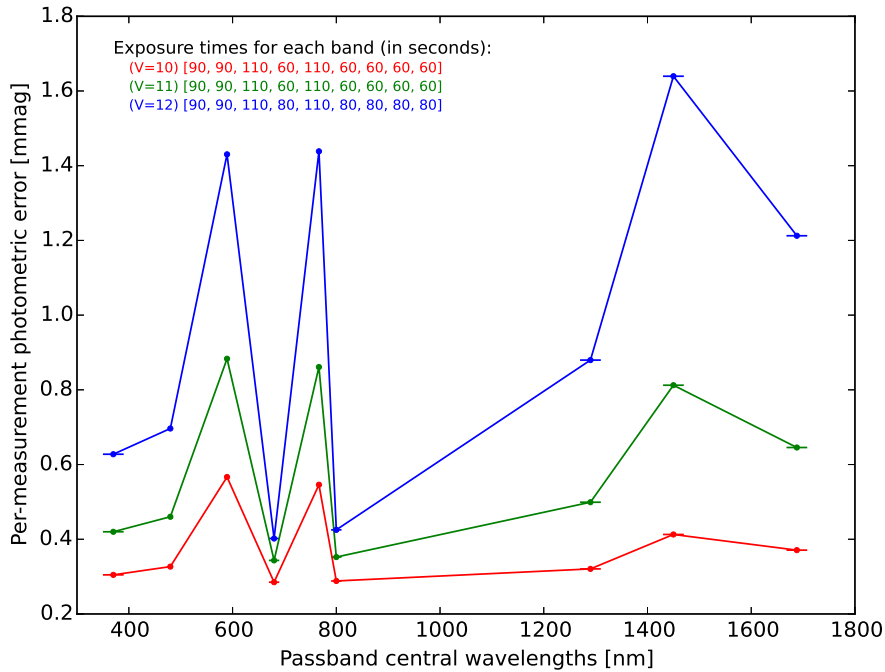} 
\caption{Expected errors per measurement (in mmag) for each of the passbands considered in this work (Sect.~\ref{subsec:OptoMechanicalSolutions}). We considered a G2V star with three different apparent magnitudes: V=10 (red dots); V=11 (green dots); V=12 (blue dots). The simulated exposure times for each passband are indicated in the overplotted text.}
\label{instsitemod}
\end{center}
\end{figure}

For the case discussed in this work we assume a 3.6~m telescope. 
We show in Fig.~\ref{instsitemod} the predicted photometric errors (in mmag) per measurement for the passbands defined in Sect.~\ref{subsec:OptoMechanicalSolutions} and the three considered apparent magnitudes. The reference exposure times, which are inset in the plot, were selected to avoid saturation and are lower than 120~seconds, that we assumed as the upper limit for our sampling strategy. 

\subsubsection{Observing Campaign Simulator}
\label{subsubsec:ObservingCampaignSimulator}

The second module of the scientific simulator aims at evaluating our capability to recover the radius variations as a function of wavelength due to the presence of chemical species in the atmosphere of a transiting planet. 
For this purpose, we injected transit signals into synthetic light curves for each passband with transit depths corresponding to the planetary radii as predicted by the atmospheric models of Fortney et al. (2010)\cite{2010ApJ...709.1396F}.
In particular, we used, as a comparative case, a transiting planet with mean radius $R_{\rm p}$=1.3~R$_\mathrm{Jup}$, orbital period $P=1.43$~days, and inclination $i=90$~deg, orbiting the primary with radius $R_\star=1$~R$_\odot$ and mass $M_\star=1$~M$_\odot$. 

For the planet we chose an atmospheric model with $T_{\rm eq}=2000$~K, surface gravity 25~$\rm m~ s^{-2}$, and without TiO and haze.
For each band we used a quadratic limb-darkening (LD hereafter) law.

We considered a time sampling between 60~s and 90~s for the nine passbands (see previous section) with a single point uncertainty between 0.3~mmag (wider bands and lower magnitude of the host star) and 1.7~mmag (narrower bands and higher stellar magnitude), and three transit events in total.

The signal reconstruction was carried out using a least-square method with the Levenberg-Marquardt algorithm and the Mandel \& Agol~(2002) \cite{mandelagol} formalism for the transit model, by fixing the two LD coefficients and the orbital period (supposed to be well known). 
 
\begin{figure}[h!]
\begin{center}
\includegraphics[scale=0.4]{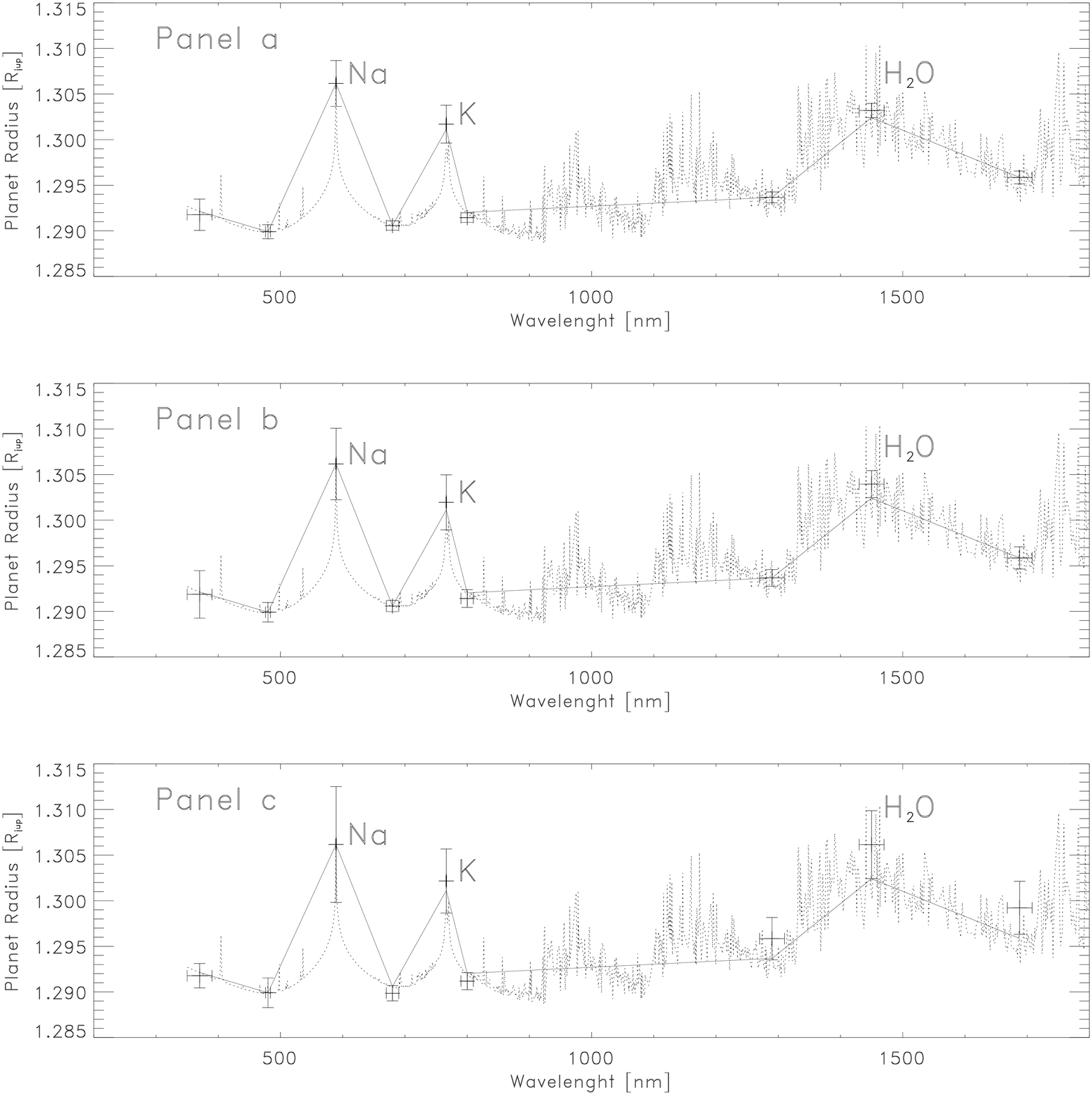} 
\vspace{0.5cm}
\caption{Planet radius [$R_{\rm Jup}$] vs wavelength [nm]. The synthetic planetary spectrum is plotted with a dotted line. Crosses with error bars are the retrieved planetary radii of the injected atmospheric absorption signals for a hot Jupiter with $T_{\rm eq}=2000$~K and $g=25~\rm m ~s^{-2}$ transiting a G2V star with three different apparent magnitudes: V=10 (panel a); V=11 (panel b); V=12 (panel c). Note the increasing uncertainties on $R_{\rm p}$ as we move to fainter stars.}
\label{fig:transitfit}
\end{center}
\end{figure}

Figure~\ref{fig:transitfit} shows the retrieved planetary radius as a function of wavelenght with overplotted the synthetic planetary spectrum (dotted lines). The three panels correspond to the three different stellar magnitudes we considered. Horizontal bars indicate the width of the nine passbands (Sect.~\ref{subsec:OptoMechanicalSolutions}) and the vertical error bars are the uncertainties on $R_{\rm p}$ as determined from the covariance matrix.

In Table~\ref{tab:snr} we report on the main results of our preliminary simulations expressed in terms of the SNR of the detection of the species Na, K, and H$_2$O. The SNR was computed as the difference between the retrieved planetary radius at the peak of the absorption features and that of the adjacent continuum by taking the uncertainties on $R_{\rm p}$ values into account.

\begin{table} [h!]
\caption{SNR for the main three atmospheric absorption bands as a function of the magnitude of the host star.} 

\label{tab:snr}
\begin{center}       
\begin{tabular}{|l|c|c|c|c|l|c|} 
\hline
& V=10 & V=11 & V=12\\
\hline
Na &  6.1  &  3.9   & 2.5 \\\hline
K &   4.9  &  3.5   & 3.2 \\\hline
H2O & 8.3  &  5.0   & 1.9 \\\hline
\end{tabular}
\end{center}
\end{table}

Additional simulations are ongoing 
to test whether a different number and/or different widths of the adopted narrow-band filters (Sect.~\ref{subsec:OptoMechanicalSolutions})
may yield more robust detections with less dedicated
telescope time per target. 
These new simulations make also use of 
more sophisticated Bayesian transit fitting
techniques (e.g., Bonomo et al. (2015)\cite{2015A&A...575A..85B}).

\subsection{Opto-mechanical solutions}
\label{subsec:OptoMechanicalSolutions}

The project of a simultaneous multi-band camera for the characterization of exoplanets with a photometrical study has already started. 
We have pointed out three main sub-project investigated in parallel:

\begin{enumerate}
\item \textbf{Chromolo - A} a radial distribution of the optical flux using dichroic filters for the wavelength separation and narrow band filters or liquid crystal filters for the observations
\item \textbf{Chromolo - B} a tree distribution of the optical flux (that means $2^{n}$ focuses). The same technique as in point 1 is used for the beam separation and filtering
\item \textbf{Chromolo - C} the most exotic solution is to study a complete optical system (i.e. a brand new telescope) that exploit the chromatic error of a reflecting surface for focusing the different wavelength in different location at the focal plane. 
\end{enumerate}

The main requirements and specifications for such an optical system are those initially defined based on the simulation work described above and they are briefly summarized here table:

\begin{itemize}
\item $370 \pm 20$ nm ,`Rayleigh Scattering'
\item $480 \pm 4$ nm 
\item $589.3 \pm 0.6$ nm , `Na Doublet'
\item $ 680 \pm 10$ nm 
\item $766.5 \pm 0.6$ nm , `K doublet'
\item $800 \pm 10$ nm 
\item $1280 \pm 20$ nm 
\item $1450 \pm 20$ nm , `H$_2$O'
\item $1685 \pm 20$ nm 
\end{itemize}

where for each band the FOV should be at least of 5'x5' up to 20'x20' with a typical PSF FWHM of 4 to 10 arcsec with a sampling around 0.15-0.6 arcsec/pixel.

\subsubsection{Radial Solution}
\label{subsubsec:Solution1}


As shown in figure \ref{fig:filters01} the fundamental idea is to design a camera that could be mounted on basically any telescope, assuring the minimal flux needed. 
The dichroic LWP (Low Wavelength Pass) or HWP (High Wavelength Pass) will split the main optical bundle separating it over a broad wavelength range. 
Before focusing on the sensor a narrow band filter or a liquid crystal tunable (hereafter LCT Filter or LCTF)filter will take place for selecting the right wavelength. 
The behaviour of the LCT Filters allows to change the band at any moment during the observation and allows a best fit to the observational requests without any HW intervention on the camera system. 
Specific choices for the dichroic scheme are being investigated. 
A concept for an innovative way of implementation of simultaneity of the measurements and filter selection is described below.

\begin{figure}[h!]
\begin{center}
\includegraphics[scale=0.45]{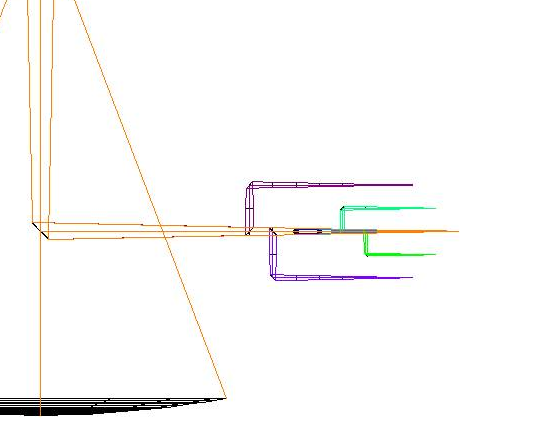} 
\caption{Schematic of the radial configuration of Chromolo camera with a proxy telescope}
\label{fig:chrom_A}
\end{center}
\end{figure}

\paragraph{Synchronous acquisition}
\label{par:SynchronousAcquisition}

As outlined in the scientific simulator section, it is crucial to obtain a strong and repeatable coherence in the frames acquisition. 
In fact only if all the observations at different wavelength  are perfectly synchronous the data obtained could be together and add a strong parameter to the planetary model.
The way to obtain this result in our multi-$\lambda$ camera is to fix the MJD (Mid Julian Date, hence the JD at half exposure time) of each exposure as contemporary for each $\lambda$-frame. 
This is the only option to have frames aligned along the time axis, thus obtaining a cubic fits where each frame slice has the same MJD. 
Otherway it is impossible to have the starting and ending time equal for all $\lambda$-frames due to the flux differences on the SED of each target.
Achieving the contemporaneity at MJD will be made possible by installing a well calibrated splitting dichroic (i.e 92\% of flux transmitted, 8\% of flux reflected) on the light path just before the focal extraction and after the wavelength filters.
On the 8\% reflected ligth path a photodiodes will be put in order to estimate the amount of the flux joining the sensor at the focal extraction. 
Having such an information will permit to fine estimate the exposure needed for the CCD imaging phase. 
A dedicated management SW will reduce all the information for the different wavelength branches of the instrument and calculate the correct MJD to be set for each run of exposure. 
Setting the MJD and knowing the exposure for each $\lambda$ it will be only a work on offsetting and delaying the start of each $\lambda$-frame in order to align the MJD of all the exposures.
The use of a photodiode is to have a wider dynamic range of functioning in order to have a wider span on the flux accepted by the instrument, hence a wider magnitude target sample could be reached.

Following the procedure step by step of the control flow over the acquisition run: 

\begin{itemize}
    \item after\_pointing
    \item for each photodiode start\_acquisition(15 s)
    \item readout(acquisition\_photodiode)
    \item calculate(exposure for each CCD)
    \item for each CCD calculate(offset\_for\_aligning\_MJD)
    \item readout(CCD\_acquisition)
\end{itemize}

\subsubsection{Xmas tree Solution}
\label{subsubsec:Solution2}


As already said the technique for separating all different wavelengths is the same as in\ref{subsubsec:Solution1}. We are exploring dichroic filters in order to split bands into different optical path. 
This solution has been investigated because with a tree separation of the wavelengths the light will pass through less optical surfaces, and so there is a fewer absorption in term of flux. 
The biggest problem of such a configuration is the strong dependency of the number of wavelength to be separated that has to be $2^{n}$ with n positive integer. 
As pointed out before we have to investigate exoplanets in 9 bands so, if it is not possible to trim out one wavelength, we must extract, with this configuration, 16 foci with 7 of them unused. 
This is a waste of flux. It is now under study the possibility to reduce to 8 the bands to be studied.

\subsubsection{Chromatic Solution}
\label{subsubsec:Solution3}

For this part of the project we have concluded the simulation about the PSF centroid position at the focal plane of an imaginary telescope. 
The way to investigate the feasibility of a strong chromatic telescope is to study the movement of the PSF centroid at the focal plane in function of some coefficients of Zernike polynomials. 
Starting from the study made for GAIA mission \cite{chroma} we already know that the most influencing terms of the 
Zernike polynomials are  6,9,14,15,19,20,21.
The simulation is built on a MCMC multiple chain for being sure about the convergence of likelihood that is estimated as a simple distance between the PSF maxima. 
A complete simulation on multiple chain will be ran soon, and for completeness not only the space of the coefficient maximizing the \textit{chromaticity} will be studied, but a full investigation for 21 parameters has been done. 
The simulation carried out has as input parameters the aperture of the optical system and the f/\#. The aperture can vary from 0.5 m to 4.0 m with a step of 50 cm. 
For the f/\# we have chosen some typical values for photometrical telescope of this range of aperture.
The MCMC allow us to trim dramatically the number of iterations and so it is a big gain in  execution time.
The MCMC simulation outlined a problem that at the beginning was not encountered, indeed a lot of couple of Zernike polynomials coefficients are complementary, hence moving one parameter leads to compensate the barycenter offset of the PSF of another one.
We decided to simulate singularly each Zernike polynomial coefficient in order to estimate the entity of the PSF barycenter displacement, regarding the center of the focal plane.

\begin{figure}[h!]
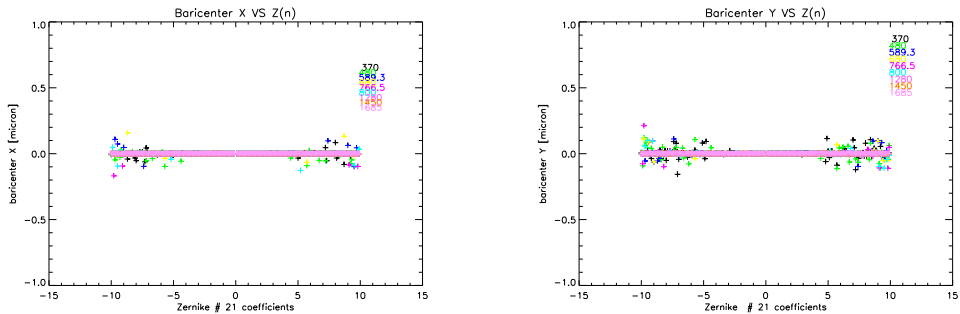

\begin{center}
\begin{subfigure}{.4\textwidth}
\includegraphics[trim=3cm 13cm 0cm 2cm,scale=0.33,page=10]{Pictures/barixy_zern21_4_12.pdf}
\caption{Baricenter along X calculated at different wavelength for a telescope with an Aperture 4.0 m and f/ \# 12.0}
\label{subfig:barix}
\end{subfigure}
\begin{subfigure}{.4\textwidth}
\includegraphics[trim=3cm 13cm 0cm 2cm,scale=0.33,page=11]{Pictures/barixy_zern21_4_12.pdf}
\caption{Baricenter along Y axis calculated at different wavelength for a telescope with an Aperture 4.0 m and f/ \# 12.0}
\end{subfigure}
\caption{Example of the Zernike parameters simulation}
\label{fig:bari}
\end{center}
\end{figure}

As showed in fig. \ref{fig:bari} the barycenter elongation, hence the chromaticity of the system is more evident at higher aperture and f-number. 
This behaviour is in average maintained for the other lower degree coefficients of the Zernike Polynomials.

\subsubsection{Liquid Crystal Tunable Filters}
\label{subsubsec:LiquidCrystalTunableFilters}

Liquid crystal tunable filters (LCTFs) are optical filters that use electronically controlled liquid 
crystal (LC) elements to transmit a selectable wavelength of light and exclude others. 
The main difference with the original Lyot filter is that the fixed wave plates are replaced by switchable liquid crystal wave plates.
LCTFs are known for enabling very high image quality and allowing relatively easy integration with regard to optical system design and software control but having lower peak transmission values in comparison with conventional fixed-wavelength optical filters due to the use of multiple polarizing elements. 
This can be mitigated in some instances by using wider bandpass designs, since a wider bandpass results in more light traveling through the filter. 
Some LCTFs are designed to tune to a limited number of fixed wavelengths such as the red, green, and blue (RGB) colors while others can be tuned in small increments over a wide range of wavelengths such as the visible or near-infrared spectrum from 400 to the current limit of 2450 nm. 
The tuning speed of LCTFs varies by manufacturer and design, but is generally several tens of milliseconds, mainly determined by the switching speed of the liquid crystal elements. 
There is a strong dependency of the tuning speed against the working temperature, lower temperature makes the viscosity higher hence an higher time to tune is necessary, this should be minimized at higher temperatures.
Implement a set of LCTFs in a multiwavelgth camera fr astronomical purposes increase the capability of the optical design to reach different wavelegth at the same focal extration exceeding the capabilties of a classical filter. 
There is also another type of solid-state tunable filter: the Acousto Optic Tunable Filter, based on the acousto-optic effect of sound waves to diffract and shift the frequency of light, but they are discarded due to their poorness in imaging quality.
LCTFs are capable of diffraction-limited imaging onto high-resolution imaging sensors and they can have working aperture sizes up to 35mm and can be placed into positions where light rays travel through the filter at angles of over 7 degrees from the normal.
LCTFs can have a long lifespan (many years) and they are lightweight, capabilities that can make possible even to design a portable instruments that could be used on a wider set of instruments.

Environmental factors that can cause degradation of filters are extended exposure to high heat and humidity, thermal and/or mechanical shock, and long-term exposure to high photonic energy such as ultraviolet light which can photobleach some of the materials used to construct the filters.
Hence the use of such LCTFs could increase the the wavelength range of observation avoiding mobile parts as filter wheel and making more narrow the filtering.
For all these reasons LCTFs are outlined as the best solution for focal filtering in our design.

\subsubsection{Dichroics study}
\label{subsubsec:DichroicStudy}

\begin{figure}[h!]
   \begin{center}
   \begin{tabular}{c}
   \includegraphics[height=6cm]{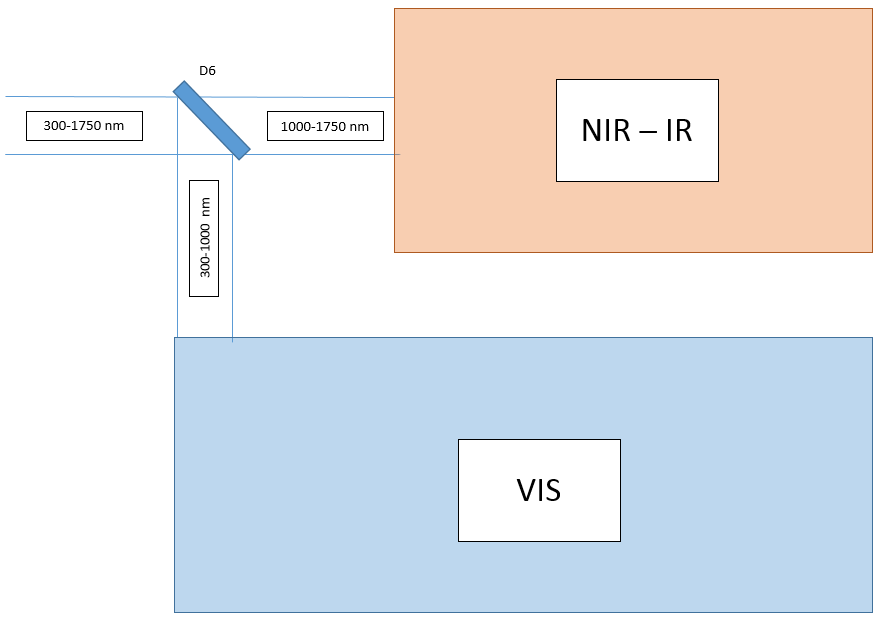}
   \end{tabular}
   \end{center}
   \caption[Filters01] 
   { \label{fig:filters01} 
Schematic division between visible and infrared branches.}
   \end{figure}

\begin{figure}[h!]
\begin{center}
\begin{subfigure}{.4\textwidth}
\includegraphics[scale=0.25]{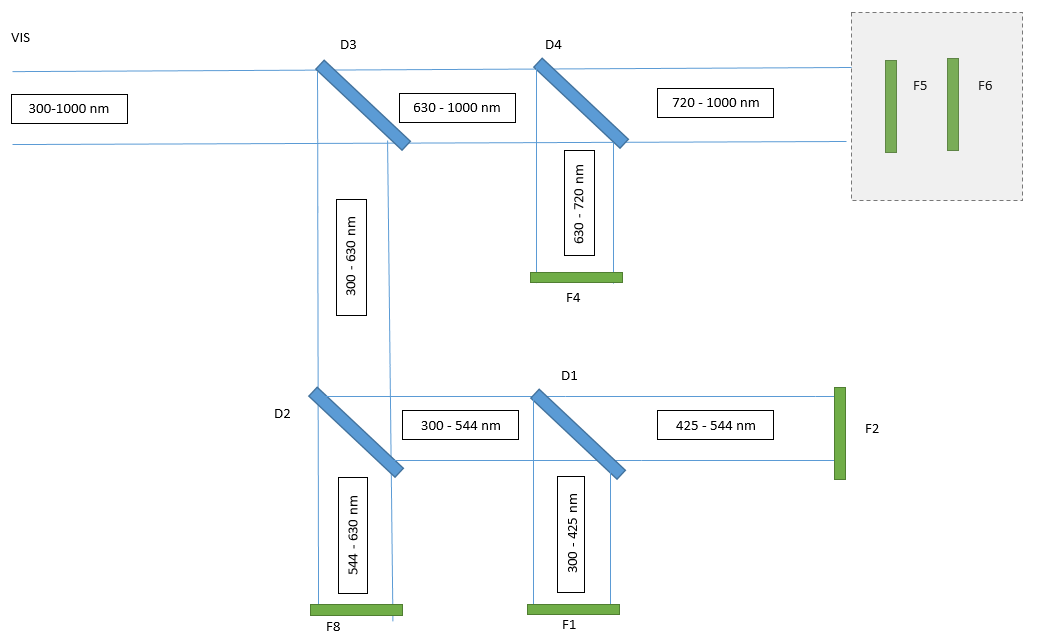}
\caption{Schematic division of the visible arm.}
\label{subfig:filters02}
\end{subfigure}
\begin{subfigure}{.4\textwidth}
\includegraphics[scale=0.285]{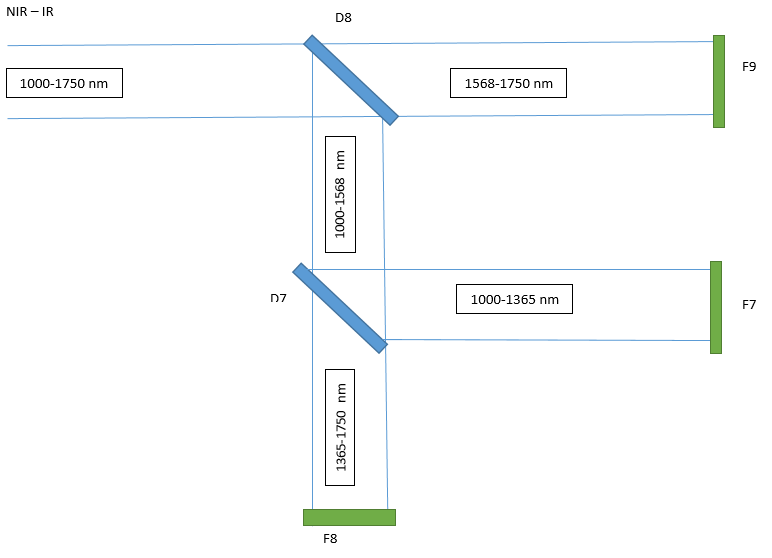}
\caption{Schematic division of the infrared arm.}
\label{subfig:filters03}
\end{subfigure}
\caption{Schematic division of the optical path of the multiband camera}
\label{fig:filters}
\end{center}
\end{figure}

In this section, we will discuss the most important aspects of the dichroics use for the wavelength separation.
We analysed and studied the case proposed in (see \ref{subsubsec:Solution2}). 
Figure \ref{fig:filters01}, \ref{subfig:filters02}, \ref{subfig:filters03}, shows the schematic principle for the filters displacement.

A first wide band dichroic filter, with an angle work of 15$^\circ$, splits the beam in two arms, the IR arm and the visible one.  Subsequently two dichroic trains produce eight channels of different wavelength, two dichroics split the beam in three IR channels and other four produce five Vis channels. 
All dichroics work at an angle of 45$^\circ$, and are long wavelength pass.  At the end of each channel a filter select the relative band indicated in \ref{subsec:OptoMechanicalSolutions}. 

In \ref{subsubsec:Solution2} it is pointed that for two channels, 766,5 $\pm$  0,6 nm and 800 $\pm$ 10 nm,  the closeness in wavelength and the narrow band requirement is severe. 
In this case, our proposal is to introduce the possibility of switching between two narrow filters placed in the same channel of slightly wider band in order to obtain the maximum efficiency for both bands.

The initial study suggests that the sharpness of the band can cause some problems to the Radial Solution too (see \ref{subsubsec:Solution1}). Furthermore the Radial Solution needs of dichroic filters that reflect with high efficiency a selected narrow band and at the same time transmit an extremely wide band with high efficiency too.
A deeper study will allow to have better information in order to understand if the Radial Solution will be feasible or not.
Table \ref{tab:Dichroics} shows all the characteristics of dichroics needed for the Xmas tree Solution, while Table \ref{tab:Filters} reports the theoretical efficiency of every channel calculated taking into account filters and dichroic filters only, without considering any other possible optical element that could be present in the beam.
Some channels are subjected to a greater number of transmission with respect to the others.
Our proposal is to have the higher efficiencies for the IR channels and for the narrow band channels.

\begin{table} [h!]
\caption{Dichroics.} 
\label{tab:Dichroics}
\begin{center}       
\begin{tabular}{|l|c|c|c|c|l|c|} 
\hline
DICHROICS & Cut Range & Transm. & Refl. Range & Transm. Range & Transition & Working \\
 & $\lambda$ [nm] &$\lambda$ [nm] & R $>$ 95\% [nm] & T $>$ 95\% [nm] & 20\% - 80\% [nm] & Angle \\
\hline
D6 (first) * & 810 $\leftrightarrow$ 1260 & 1000 & 350 - 810 & 1260 - 1710 & 150 (925 - 1075) & 15$^\circ$\\\hline
D8 (second - IR) & 1470 $\leftrightarrow$ 1665 & 1570 & 1260 - 1470 & 1660 - 1705 & 50 (1545 - 1595) & 45$^\circ$\\\hline
D3 (second -VIS) & 590 $\leftrightarrow$ 670 & 630 & 350 - 590 & 670 - 810 & 20 (620 - 640) & 45$^\circ$\\\hline
D7 (third - IR) & 1300 $\leftrightarrow$ 1430 & 1365 & 1260 - 1300 & 1430 - 1470 & 30 (1350 - 1380) & 45$^\circ$\\\hline
D4 (third - VIS) & 690 $\leftrightarrow$ 760 & 730 & 670 - 690 & 760 - 810 & 20 (720 - 740) & 45$^\circ$\\\hline
D2 (third - VIS) & 500 $\leftrightarrow$ 588 & 545 & 350 - 500 & 588 - 590 & 20 (535 - 555) & 45$^\circ$\\\hline
D1 (fourth - VIS) & 390 $\leftrightarrow$ 460 & 425 & 350 - 390 & 460 - 500 & 20 (415 - 435 ) & 45$^\circ$\\\hline
\end{tabular}
\end{center}
\end{table}

\begin{table} [h!]
\caption{Filters.} 
\label{tab:Filters}
\begin{center}       
\begin{tabular}{|l|c|c|c|c|l|c|} 
\hline
FILTERS & Range & Number of & Efficiency \\
 & [nm] & passages & (theoric) \\
\hline
F1 & 350 $\leftrightarrow$ 390 & 4 & 70\% \\\hline
F2 & 460 $\leftrightarrow$ 500 & 4 & 70\% \\\hline
F3 & 589,3 $\pm$ 0,6 & 3 & 77\% \\\hline
F4 & 670 $\leftrightarrow$ 690 & 3 & 77\% \\\hline
F5 & 766,5 $\pm$ 0,6 & 3 & 77\% \\\hline
F6 & 790 $\leftrightarrow$ 810 & 3 & 77\% \\\hline
F7 & 1260 $\leftrightarrow$ 1300  & 3 & 77\% \\\hline
F8 & 1430 $\leftrightarrow$ 1470 & 3 & 77\% \\\hline
F9 & 1665 $\leftrightarrow$ 1705 & 2 & 80\% \\\hline
\end{tabular}
\end{center}
\end{table}

\section{Summary \& Discussion}
\label{sec:summary}

The development of instrumentation specifically designed for the purpose of exoplanets' atmospheres characterization will allow in the future to move from the 
realm of exploratory work to that of systematic studies enabling the {\it repeatability} of the measurements. Similarly to the case of planet detection in the first place, 
the best results will be obtained by the implementation of a multi-technique approach, exploiting the potential of low- to high-resolution spectroscopy and very high-precision 
photometry, both from the ground and in space, and over a broad range of wavelength. 

We have presented here initial results of an ongoing feasibility study for the prototype of a versatile multi-band imaging
system that exploits the choice of $a)$ specifically selected narrow-band filters and b) new ideas for the execution of
high-precision simultaneous measurements at VIS $\&$ NIR wavelengths for the systematic characterization of extrasolar planets' atmospheres.

The study has been setup with a two-tiered approach focused on $1)$ the development of a scientific simulator aimed at quantifying the performance 
in the retrieval of the fundamental physical quantity ($R_p$ vs. $\lambda$ ) as a function of instrument parameters, choices for the narrow-band filters, 
and details of proxies for observing campaigns. The simulator has allowed us to establish the fundamental system requirements driving $2)$ the investigation 
of a number of opto-mechanical solutions for the prototype.

To understand the diversity of the atmospheres of hot Jupiters, that is how their compositions and circulations 
are related to planet formation, migration, and interactions with the host stars, 
we need an ever growing number of well-characterised planetary atmospheres. Our prototype may increase to a greater extent the number of well-studied atmospheres of giant planets 
around solar-like stars and of some Neptunes orbiting late K and M dwarfs. 
Only this way we will be able to search for correlations of the properties of exoplanetary atmospheres 
with both stellar and planetary parameters and evolution histories, in order to build a theoretical framework
that may be able to explain their diversity. 

Upon identification of the final instrument design, such an imaging system will become a very important addition to the lot of existing and planned instruments devoted 
to exoplanets' atmospheric studies, particularly when seen as fundamental complement to lower-resolution spectroscopic measurements from space at IR wavelengths (with e.g. JSWT,  
ESA's M4 proposed mission ARIEL), as well as higher-resolution ground-based spectroscopy at VIS and NIR wavelengths carried out with e.g., HARPS and its NIR extension NIRPS, the 
GIARPS (GIANO+HARPS-N) facility, ESPRESSO@VLT, and HIRES@E-ELT.


\acknowledgments
Jean Marc Christille, Mario Damasso,Francesco Borsa and Paolo Giacobbe acknowledge support from INAF through the "Progetti Premiali" funding scheme of the Italian Ministry of Education, University, and Research.       
 
\nocite{*}
\bibliography{SPIE_SIOUX} 

\begin{thebibliography}{10}

\bibitem{2010Sci...330..653H}
{Howard}, A.~W., {Marcy}, G.~W., {Johnson}, J.~A., {Fischer}, D.~A., {Wright},
  J.~T., {Isaacson}, H., {Valenti}, J.~A., {Anderson}, J., {Lin}, D.~N.~C., and
  {Ida}, S., ``{The Occurrence and Mass Distribution of Close-in Super-Earths,
  Neptunes, and Jupiters},'' {\em Science}~{\bf 330},  653 (Oct. 2010).

\bibitem{2013ApJ...766...81F}
{Fressin}, F., {Torres}, G., {Charbonneau}, D., {Bryson}, S.~T.,
  {Christiansen}, J., {Dressing}, C.~D., {Jenkins}, J.~M., {Walkowicz}, L.~M.,
  and {Batalha}, N.~M., ``{The False Positive Rate of Kepler and the Occurrence
  of Planets},'' {\em The Astrophysical Journal}~{\bf 766},  81 (Apr. 2013).

\bibitem{2015ApJ...814..130M}
{Mulders}, G.~D., {Pascucci}, I., and {Apai}, D., ``{An Increase in the Mass of
  Planetary Systems around Lower-mass Stars},'' {\em The Astrophysical
  Journal}~{\bf 814},  130 (Dec. 2015).

\bibitem{2015ApJ...807...45D}
{Dressing}, C.~D. and {Charbonneau}, D., ``{The Occurrence of Potentially
  Habitable Planets Orbiting M Dwarfs Estimated from the Full Kepler Dataset
  and an Empirical Measurement of the Detection Sensitivity},'' {\em The
  Astrophysical Journal}~{\bf 807},  45 (July 2015).

\bibitem{2016A&A...587A..64S}
{Santerne}, A., {Moutou}, C., {Tsantaki}, M., {Bouchy}, F., {H{\'e}brard}, G.,
  {Adibekyan}, V., {Almenara}, J.-M., {Amard}, L., {Barros}, S.~C.~C.,
  {Boisse}, I., {Bonomo}, A.~S., {Bruno}, G., {Courcol}, B., {Deleuil}, M.,
  {Demangeon}, O., {D{\'{\i}}az}, R.~F., {Guillot}, T., {Havel}, M.,
  {Montagnier}, G., {Rajpurohit}, A.~S., {Rey}, J., and {Santos}, N.~C.,
  ``{SOPHIE velocimetry of Kepler transit candidates. XVII. The physical
  properties of giant exoplanets within 400 days of period},'' {\em Astronomy
  and Astrophysics}~{\bf 587},  A64 (Mar. 2016).

\bibitem{2010ARA&A..48..631S}
{Seager}, S. and {Deming}, D., ``{Exoplanet Atmospheres},'' {\em Annual Review
  of Astronomy and Astrophysics}~{\bf 48},  631--672 (Sept. 2010).

\bibitem{2014Natur.513..345B}
{Burrows}, A.~S., ``{Highlights in the study of exoplanet atmospheres},'' {\em
  Nature}~{\bf 513},  345--352 (Sept. 2014).

\bibitem{2015PASP..127..941C}
{Crossfield}, I.~J.~M., ``{Observations of Exoplanet Atmospheres},'' {\em
  Publications of the Astronomical Society of Pacific}~{\bf 127},  941--960
  (Oct. 2015).

\bibitem{2010ApJ...709.1396F}
{Fortney}, J.~J., {Shabram}, M., {Showman}, A.~P., {Lian}, Y., {Freedman},
  R.~S., {Marley}, M.~S., and {Lewis}, N.~K., ``{Transmission Spectra of
  Three-Dimensional Hot Jupiter Model Atmospheres},'' {\em The Astrophysical
  Journal}~{\bf 709},  1396--1406 (Feb. 2010).

\bibitem{2014prpl.conf..763B}
{Baraffe}, I., {Chabrier}, G., {Fortney}, J., and {Sotin}, C., ``{Planetary
  Internal Structures},'' {\em Protostars and Planets VI} ,  763--786 (2014).

\bibitem{2016arXiv160104761F}
{Fischer}, P.~D., {Knutson}, H.~A., {Sing}, D.~K., {Henry}, G.~W.,
  {Williamson}, M.~W., {Fortney}, J.~J., {Burrows}, A.~S., {Kataria}, T.,
  {Nikolov}, N., {Showman}, A.~P., {Ballester}, G.~E., {D{\'e}sert}, J.-M.,
  {Aigrain}, S., {Deming}, D., {Lecavelier des Etangs}, A., and {Vidal-Madjar},
  A., ``{HST hot-Jupiter transmission spectral survey: Clear skies for cool
  Saturn WASP-39b},'' {\em ArXiv e-prints}  (Jan. 2016).

\bibitem{2013MNRAS.432.2917P}
{Pont}, F., {Sing}, D.~K., {Gibson}, N.~P., {Aigrain}, S., {Henry}, G., and
  {Husnoo}, N., ``{The prevalence of dust on the exoplanet HD 189733b from
  Hubble and Spitzer observations},'' {\em Monthly Notices of the Royal
  Astronomical Society}~{\bf 432},  2917--2944 (July 2013).

\bibitem{2015MNRAS.447..463N}
{Nikolov}, N., {Sing}, D.~K., {Burrows}, A.~S., {Fortney}, J.~J., {Henry},
  G.~W., {Pont}, F., {Ballester}, G.~E., {Aigrain}, S., {Wilson}, P.~A.,
  {Huitson}, C.~M., {Gibson}, N.~P., {D{\'e}sert}, J.-M., {Etangs}, A.~L.~d.,
  {Showman}, A.~P., {Vidal-Madjar}, A., {Wakeford}, H.~R., and {Zahnle}, K.,
  ``{HST hot-Jupiter transmission spectral survey: haze in the atmosphere of
  WASP-6b},'' {\em Monthly Notices of the Royal Astronomical Society}~{\bf
  447},  463--478 (Feb. 2015).

\bibitem{2015MNRAS.446.2428S}
{Sing}, D.~K., {Wakeford}, H.~R., {Showman}, A.~P., {Nikolov}, N., {Fortney},
  J.~J., {Burrows}, A.~S., {Ballester}, G.~E., {Deming}, D., {Aigrain}, S.,
  {D{\'e}sert}, J.-M., {Gibson}, N.~P., {Henry}, G.~W., {Knutson}, H.,
  {Lecavelier des Etangs}, A., {Pont}, F., {Vidal-Madjar}, A., {Williamson},
  M.~W., and {Wilson}, P.~A., ``{HST hot-Jupiter transmission spectral survey:
  detection of potassium in WASP-31b along with a cloud deck and Rayleigh
  scattering},'' {\em Monthly Notices of the Royal Astronomical Society}~{\bf
  446},  2428--2443 (Jan. 2015).

\bibitem{2014ApJ...793L..27K}
{Kreidberg}, L., {Bean}, J.~L., {D{\'e}sert}, J.-M., {Line}, M.~R., {Fortney},
  J.~J., {Madhusudhan}, N., {Stevenson}, K.~B., {Showman}, A.~P.,
  {Charbonneau}, D., {McCullough}, P.~R., {Seager}, S., {Burrows}, A., {Henry},
  G.~W., {Williamson}, M., {Kataria}, T., and {Homeier}, D., ``{A Precise Water
  Abundance Measurement for the Hot Jupiter WASP-43b},'' {\em The Astrophysical
  Journal}~{\bf 793},  L27 (Oct. 2014).

\bibitem{2014ApJ...791L...9M}
{Madhusudhan}, N., {Crouzet}, N., {McCullough}, P.~R., {Deming}, D., and
  {Hedges}, C., ``{H$_{2}$O Abundances in the Atmospheres of Three Hot
  Jupiters},'' {\em The Astrophysical Journal}~{\bf 791},  L9 (Aug. 2014).

\bibitem{2016Natur.529...59S}
{Sing}, D.~K., {Fortney}, J.~J., {Nikolov}, N., {Wakeford}, H.~R., {Kataria},
  T., {Evans}, T.~M., {Aigrain}, S., {Ballester}, G.~E., {Burrows}, A.~S.,
  {Deming}, D., {D{\'e}sert}, J.-M., {Gibson}, N.~P., {Henry}, G.~W.,
  {Huitson}, C.~M., {Knutson}, H.~A., {Etangs}, A.~L.~D., {Pont}, F.,
  {Showman}, A.~P., {Vidal-Madjar}, A., {Williamson}, M.~H., and {Wilson},
  P.~A., ``{A continuum from clear to cloudy hot-Jupiter exoplanets without
  primordial water depletion},'' {\em Nature}~{\bf 529},  59--62 (Jan. 2016).

\bibitem{2008ApJ...686..667S}
{Sing}, D.~K., {Vidal-Madjar}, A., {Lecavelier des Etangs}, A., {D{\'e}sert},
  J.-M., {Ballester}, G., and {Ehrenreich}, D., ``{Determining Atmospheric
  Conditions at the Terminator of the Hot Jupiter HD 209458b},'' {\em The
  Astrophysical Journal}~{\bf 686},  667--673 (Oct. 2008).

\bibitem{2014Sci...346..838S}
{Stevenson}, K.~B., {D{\'e}sert}, J.-M., {Line}, M.~R., {Bean}, J.~L.,
  {Fortney}, J.~J., {Showman}, A.~P., {Kataria}, T., {Kreidberg}, L.,
  {McCullough}, P.~R., {Henry}, G.~W., {Charbonneau}, D., {Burrows}, A.,
  {Seager}, S., {Madhusudhan}, N., {Williamson}, M.~H., and {Homeier}, D.,
  ``{Thermal structure of an exoplanet atmosphere from phase-resolved emission
  spectroscopy},'' {\em Science}~{\bf 346},  838--841 (Nov. 2014).

\bibitem{2008ApJ...678.1419F}
{Fortney}, J.~J., {Lodders}, K., {Marley}, M.~S., and {Freedman}, R.~S., ``{A
  Unified Theory for the Atmospheres of the Hot and Very Hot Jupiters: Two
  Classes of Irradiated Atmospheres},'' {\em The Astrophysical Journal}~{\bf
  678},  1419--1435 (May 2008).

\bibitem{2014Natur.505...69K}
{Kreidberg}, L., {Bean}, J.~L., {D{\'e}sert}, J.-M., {Benneke}, B., {Deming},
  D., {Stevenson}, K.~B., {Seager}, S., {Berta-Thompson}, Z., {Seifahrt}, A.,
  and {Homeier}, D., ``{Clouds in the atmosphere of the super-Earth exoplanet
  GJ1214b},'' {\em Nature}~{\bf 505},  69--72 (Jan. 2014).

\bibitem{2014Natur.505...66K}
{Knutson}, H.~A., {Benneke}, B., {Deming}, D., and {Homeier}, D., ``{A
  featureless transmission spectrum for the Neptune-mass exoplanet GJ436b},''
  {\em Nature}~{\bf 505},  66--68 (Jan. 2014).

\bibitem{2012A&A...539A.140B}
{Ballerini}, P., {Micela}, G., {Lanza}, A.~F., and {Pagano}, I.,
  ``{Multiwavelength flux variations induced by stellar magnetic activity:
  effects on planetary transits},'' {\em Astronomy and Astrophysics}~{\bf 539},
   A140 (Mar. 2012).

\bibitem{2008PASP..120..405G}
{Greiner}, J., {Bornemann}, W., {Clemens}, C., {Deuter}, M., {Hasinger}, G.,
  {Honsberg}, M., {Huber}, H., {Huber}, S., {Krauss}, M., {Kr{\"u}hler}, T.,
  {K{\"u}pc{\"u} Yolda{\c s}}, A., {Mayer-Hasselwander}, H., {Mican}, B.,
  {Primak}, N., {Schrey}, F., {Steiner}, I., {Szokoly}, G., {Th{\"o}ne}, C.~C.,
  {Yolda{\c s}}, A., {Klose}, S., {Laux}, U., and {Winkler}, J., ``{GROND a
  7-Channel Imager},'' {\em Publications of the Astronomical Society of
  Pacific}~{\bf 120},  405--424 (Apr. 2008).

\bibitem{2013MNRAS.430.2932M}
{Mancini}, L., {Nikolov}, N., {Southworth}, J., {Chen}, G., {Fortney}, J.~J.,
  {Tregloan-Reed}, J., {Ciceri}, S., {van Boekel}, R., and {Henning}, T.,
  ``{Physical properties of the WASP-44 planetary system from simultaneous
  multi-colour photometry},'' {\em Monthly Notices of the Royal Astronomical
  Society}~{\bf 430},  2932--2942 (Apr. 2013).

\bibitem{2014MNRAS.443.2391M}
{Mancini}, L., {Southworth}, J., {Ciceri}, S., {Tregloan-Reed}, J.,
  {Crossfield}, I., {Nikolov}, N., {Bruni}, I., {Zambelli}, R., and {Henning},
  T., ``{Physical properties, star-spot activity, orbital obliquity and
  transmission spectrum of the Qatar-2 planetary system from multicolour
  photometry},'' {\em Monthly Notices of the Royal Astronomical Society}~{\bf
  443},  2391--2409 (Sept. 2014).

\bibitem{2009MNRAS.396.1023S}
{Southworth}, J., {Hinse}, T.~C., {J{\o}rgensen}, U.~G., {Dominik}, M.,
  {Ricci}, D., {Burgdorf}, M.~J., {Hornstrup}, A., {Wheatley}, P.~J.,
  {Anguita}, T., {Bozza}, V., {Novati}, S.~C., {Harps{\o}e}, K.,
  {Kj{\ae}rgaard}, P., {Liebig}, C., {Mancini}, L., {Masi}, G., {Mathiasen},
  M., {Rahvar}, S., {Scarpetta}, G., {Snodgrass}, C., {Surdej}, J.,
  {Th{\"o}ne}, C.~C., and {Zub}, M., ``{High-precision photometry by telescope
  defocusing - I. The transiting planetary system WASP-5},'' {\em Monthly
  Notices of the Royal Astronomical Society}~{\bf 396},  1023--1031 (June
  2009).

\bibitem{mandelagol}
{Mandel}, K. and {Agol}, E., ``{Analytic Light Curves for Planetary Transit
  Searches},'' {\em The Astrophysical Journal Letters}~{\bf 580},  L171--L175
  (Dec. 2002).

\bibitem{2015A&A...575A..85B}
{Bonomo}, A.~S., {Sozzetti}, A., {Santerne}, A., {Deleuil}, M., {Almenara},
  J.-M., {Bruno}, G., {D{\'{\i}}az}, R.~F., {H{\'e}brard}, G., and {Moutou},
  C., ``{Improved parameters of seven Kepler giant companions characterized
  with SOPHIE and HARPS-N},'' {\em Astronomy and Astrophysics}~{\bf 575},  A85
  (Mar. 2015).

\bibitem{chroma}
Busonero, D., Gai, M., Gardiol, D., Lattanzi, M.~G., and Loreggia, D.,
  ``Chromaticity in all-reflective telescopes for astrometry,'' {\em Astronomy
  and Astrophysics}~{\bf 449},  827--836 (2006).

\bibitem{agaragar}
Tresoldi, D., Felletti, R., Bianco, A., Conconi, P., Caprio, V.~D., Crimi, G.,
  Molinari, E., Riva, A., Riva, M., Span\'o, P., Tintori, M., Toso, G., and
  Zerbi, F.~M., ``Agar-agar: A high-efficiency narrow-band imager for
  elts...,'' in [{\em Ground-based and Airborne Instrumentation for
  Astronomy}{\nolinebreak\hspace{0.1em}]},  McLean, I.~S., ed., {\em Proc.
  SPIE} {\bf 6269} (2006).

\bibitem{multiband}
Riva, A. and Span\'o, P., ``Compact multi-band visible camera for 1m-class fast
  telescopes,'' in [{\em Ground-based and Airborne Instrumentation for
  Astronomy II}{\nolinebreak\hspace{0.1em}]},  McLean, I.~S., ed., {\em Proc.
  SPIE} {\bf 7014} (2008).

\bibitem{octocam}
de~Ugarte~Postigo, A., Gorosabel, J., Span\'o, P., Riva, M., Rabaza, O.,
  de~Caprio, V., Cuniffe, R., Kubanek, P., Riva, A., Jelinek, M., I.Andersen,
  M., Castro-Tirado, A.~J., Zerbi, F.~M., and Fernandez-Soto, A., ``Octocam: A
  fast multichannel imager and spectrograph for the 10.4m gtc,'' in [{\em
  Ground-based and Airborne Instrumentation for Astronomy
  III}{\nolinebreak\hspace{0.1em}]},  McLean, I.~S., ed., {\em Proc. SPIE} {\bf
  7735} (2010).

\bibitem{bovis}
Riva, A., Jelinek, M., Cunniffe, R., V\'itek, S., Castro-Tirado, A.~J., Riva,
  M., and Zerbi, F.~M., ``Bovis, the visible eye of bootes-ir *,'' in [{\em
  Second Workshop on Robotic Autonomous
  Observatories}{\nolinebreak\hspace{0.1em}]},  Guziy, S., ed., {\em ASI
  Conference Series} {\bf 7},  25--32 (2012).

\bibitem{ros2}
Molinari, E., Covino, S., Crimi, G., D'Alessio, F., Incorvaia, S., Fugazza, D.,
  Span\'o, P., Toso, G., Tresoldi, D., and Vitali, F., ``Ros2, a multichannel
  vision for the robotic rem telescope.,'' in [{\em Ground-based and Airborne
  Instrumentation for Astronomy V}{\nolinebreak\hspace{0.1em}]},  McLean,
  I.~S., ed., {\em Proc. SPIE} {\bf 9147} (2014).

\bibitem{muscat}
Narita, N., Fukui, A., Kusakabe, N., Onitsuka, M., Ryu, T., Yanagisawa, K.,
  Izumiura, H., Tamura, M., and Yamamuro, T., ``Muscat: a multicolor
  simultaneous camera for studying atmospheres of transiting exoplanets,'' {\em
  Journal of Astronomical Telescopes, Instruments, and Systems}~{\bf 1(4)}
  (2015).

\bibitem{astrocomp}
Bendek, E.~A., Guyon, O., Ammond, S.~M., and Belikov, R., ``Laboratory
  demonstration of astrometric compensation using a diffractive pupil,'' {\em
  Publications of the Astronomical Society of Pacific}~{\bf 125},  1212--1225
  (2013).

\bibitem{bircam}
Riva, A., Conconi, P., Castro-Tirado, A.~J., Zerbi, F., Cunniffe, R., Jelinek,
  M., and Vitek, S., ``Design, manufacturing, and commissioning of bircam
  (bootes infrared camera),'' {\em Advances in Astronomy}~{\bf 2010},  8
  (2009).

\bibitem{grond}
Greiner, J., Bornemann, W., Clemens, C., Deuter, M., Hasinger, G., Honsberg,
  M., Huber, H., Huber, S., Krauss, M., Kruhler, T., Yoldas, A.~K.,
  Mayer-Hasselwander, H., Mican, B., Primak, N., Schrey, F., Steiner, I.,
  Szokoly, G., Thone, C.~C., and Yoldas, A., ``Grond a 7-channel imager,'' {\em
  Advances in Astronomy}~{\bf 2010},  8 (2009).

\bibitem{2012MNRAS.422.3099S}
{Southworth}, J., {Mancini}, L., {Maxted}, P.~F.~L., {Bruni}, I.,
  {Tregloan-Reed}, J., {Barbieri}, M., {Ruocco}, N., and {Wheatley}, P.~J.,
  ``{Physical properties and radius variations in the HAT-P-5 planetary system
  from simultaneous four-colour photometry},'' {\em \mnras}~{\bf 422},
  3099--3106 (June 2012).

\end{thebibliography}
\bibliographystyle{spiebib} 

\end{document}